\documentclass[showpacs,amsmath,amssymb,pra,superscriptaddress,nofootinbib]{revtex4}

\usepackage{graphicx}

\usepackage{amsmath}

\newcommand{\bra}[1]{\mbox{$\left\langle #1 \right|$}}
\newcommand{\ket}[1]{\mbox{$\left| #1 \right\rangle$}}
\newcommand{\braket}[2]{\mbox{$\left\langle #1 | #2 \right\rangle$}}

\begin{document}

\title{Source attack of decoy-state quantum key distribution using phase information}

\author{Yan-Lin Tang}
\affiliation{Hefei National Laboratory for Physical Sciences at Microscale and Department of Modern Physics, University of Science and Technology of China, Hefei, Anhui, China}

\author{Hua-Lei Yin}
\affiliation{Hefei National Laboratory for Physical Sciences at Microscale and Department of Modern Physics, University of Science and Technology of China, Hefei, Anhui, China}

\author{Xiongfeng Ma}
\email{xma@tsinghua.edu.cn}
\affiliation{Center for Quantum Information, Institute for Interdisciplinary Information Sciences, Tsinghua University, Beijing, China}

\author{Chi-Hang Fred Fung}
\email{chffung@hku.hk}
\affiliation{Department of Physics and Center of Theoretical and Computational Physics, University of Hong Kong, Pokfulam Road, Hong Kong}

\author{Yang Liu}
\affiliation{Hefei National Laboratory for Physical Sciences at Microscale and Department of Modern Physics, University of Science and Technology of China, Hefei, Anhui, China}

\author{Hai-Lin Yong}
\affiliation{Hefei National Laboratory for Physical Sciences at Microscale and Department of Modern Physics, University of Science and Technology of China, Hefei, Anhui, China}

\author{Teng-Yun Chen}
\email{tychen@ustc.edu.cn}
\affiliation{Hefei National Laboratory for Physical Sciences at Microscale and Department of Modern Physics, University of Science and Technology of China, Hefei, Anhui, China}

\author{Cheng-Zhi Peng}

\affiliation{Hefei National Laboratory for Physical Sciences at Microscale and Department of Modern Physics, University of Science and Technology of China, Hefei, Anhui, China}

\author{Zeng-Bing Chen}
\affiliation{Hefei National Laboratory for Physical Sciences at Microscale and Department of Modern Physics, University of Science and Technology of China, Hefei, Anhui, China}

\author{Jian-Wei Pan}
\email{pan@ustc.edu.cn}
\affiliation{Hefei National Laboratory for Physical Sciences at Microscale and Department of Modern Physics, University of Science and Technology of China, Hefei, Anhui, China}

\begin{abstract}
Quantum key distribution (QKD) utilizes the laws of quantum mechanics to achieve information-theoretically secure key generation. This field is now approaching the stage of commercialization, but many practical QKD systems still suffer from security loopholes due to imperfect devices. In fact, practical attacks have successfully been demonstrated. Fortunately, most of them only exploit detection-side loopholes which are now closed by the recent idea of measurement-device-independent QKD. On the other hand, little attention is paid to the source which may still leave QKD systems insecure. In this work, we propose and demonstrate an attack that exploits a source-side loophole existing in qubit-based QKD systems using a weak coherent state source and decoy states. Specifically, by implementing a linear-optics unambiguous-state-discrimination measurement, we show that the security of a system without phase randomization --- which is a step assumed in conventional security analyses but sometimes neglected in practice --- can be compromised. We conclude that implementing phase randomization is essential to the security of decoy-state QKD systems under current security analyses.
\end{abstract}


\maketitle

\section{Introduction}

Quantum key distribution (QKD) aims at offering information-theoretical security for secret key expansion \cite{Bennett:BB84:1984,Ekert:QKD:1991} that is guaranteed by quantum mechanics. Commercial QKD systems have emerged on the market and are now under rapid development. Despite theoretical security proofs, various quantum hacking strategies targeting practical QKD systems have been proposed, with some of them demonstrated in experiments. These attacks exploit certain imperfections in the devices used to build QKD systems. Except for the phase-remapping attack \cite{Fung:Remap:07,Xu:PhaseRemap:2010}, which aims at the source of plug-and-play QKD systems, until now, most practical attacks have been launched on the detection side of the QKD system, including the fake-state attack \cite{MAS_Eff_06,Makarov:Fake:08}, the time-shift attack \cite{Qi:TimeShift:2007,Zhao:TimeshiftExp:2008}, and the detector-blinding attack \cite{Lydersen:Hacking:2010,Gerhardt:Blind:2011}.

In order to achieve security when imperfect (untrusted) devices are present, QKD schemes \cite{MayersYao_98,Acin:DeviceIn:07} that are fully device independent (without assumptions on either the detector or the source) have been proposed. However, these schemes suffer from their unrealistic requirement for a high transmission efficiency (with the lowest to date being 75\% \cite{Lucamarini:DIQKD:2012}), which limits their use in practice. Recently, the newly proposed measurement-device-independent QKD (MDI-QKD) scheme \cite{Lo:MIQKD:2012}, whose security does not rely on any assumptions on the detection system, can defeat all aforementioned detection-side attacks. On the source side, the security proof of the MDI-QKD scheme relies on a trusted-source scenario, whose security concern is relatively less explored. Besides, many other security proofs \cite{Lo:Decoy:2005,LoPreskill:NonRan:2007} also rely on stringent assumptions on the source side. Any deviation from these assumptions may lead to loopholes that can be exploited for eavesdropping.

The use of different sources directly affects the security of QKD systems.
Systems implementing the popular Bennett-Brassard 1984 (BB84) protocol \cite{Bennett:BB84:1984} often use a weak coherent state (WCS) source instead of a single-photon source for the transmission of quantum states. Fortunately, such a substitution of the source is safe but its security proof \cite{GLLP:2004,Lo:Decoy:2005} assumes that the phase of the source is randomized, without which the security would be weakened \cite{Dusek:USD:2000}. Later security proof by Lo and Preskill \cite{LoPreskill:NonRan:2007} eliminated the need for phase randomization, but with the performance substantially diminished.

Even though the security of the BB84 protocol with WCS is proven, there is a substantial performance gap between it and the case of a single-photon source. A significant achievement was made with the proposal of the decoy-state technique \cite{Hwang:Decoy:2003,Lo:Decoy:2005,Wang:Decoy:2005}, which greatly improves the performance of the BB84 protocol with WCS, and thus decoy-state BB84 with WCS has become one of the most popular schemes for practical implementation. Again, phase randomization is assumed in the current security proof \cite{Lo:Decoy:2005}, and
%Among the various QKD schemes proposed, decoy-state BB84 protocol \cite{Lo:Decoy:2005} with WCS source, has become one of the most popular schemes for practical implementation \cite{Hwang:Decoy:2003}.
%While
a security analysis for decoy-state QKD without phase randomization
%the security analysis for this kind of system without phase randomization
is not yet available \cite{LoPreskill:NonRan:2007}. This fact can easily be overlooked and QKD system designers often neglect the implementation of phase randomization without realizing the danger of opening up a security loophole. Indeed, we experimentally demonstrate that this is a major security loophole. We propose and demonstrate an attack on the source part of a decoy-state QKD system with WCS when phase randomization is not implemented. By using a combination of an unambiguous-state-discrimination (USD) measurement and a photon-number-splitting (PNS) attack \cite{BLMS:PNS:2000}, we show that the final key generated by the non-phase-randomized system can be compromised.
%and thus phase randomization of the source is necessary in order to achieve security in decoy-state BB84 with WCS using current postprocessing \cite{MXF:Practical:2005}.

\section{Hacking strategy}

The essence of our hacking strategy is as follows. Since Alice prepares her pulses without phase randomization, we may assume that Eve knows
%the phase information of every transmitted pulse
the overall phase of every transmitted state by Alice in the worst-case scenario. Then, from Eve's point of view,
%the quantum signals
the states
sent by Alice are drawn from an ensemble of pure states (corresponding to the signal and decoy states).
%in pure states\footnote{Strictly speaking, they are not pure states due to the fact that Eve does not know the qubit encoded by Alice. Details will be discussed in Methods.}.
Thus, quantum mechanics allows Eve to distinguish between the signal and decoy states by a USD measurement. In this way, one of the foundations of the security proof of decoy-state QKD, the photon number channel model \cite{Lo:Decoy:2005}, is violated.

\begin{figure}[tbh]
\centering \resizebox{12cm}{!}{\includegraphics{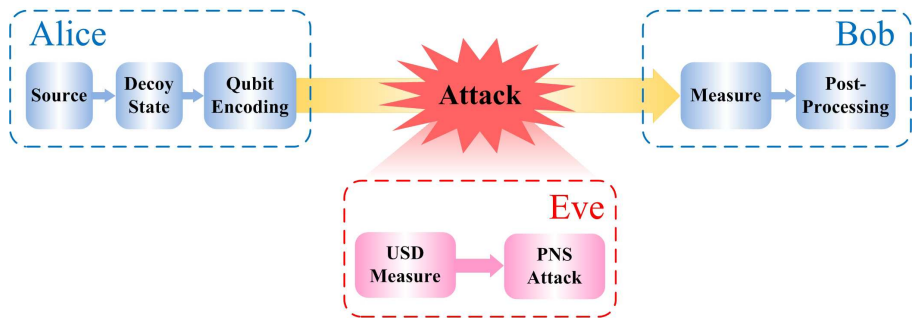}}
\caption{(Color online) Schematic diagram of the USD+PNS attack on a decoy-state QKD system without phase randomization. Eve intercepts the quantum states sent by Alice, and then performs a USD measurement to distinguish the signal/decoy state succeeded by a PNS attack. Conditioned on her results of signal/decoy information and photon numbers, she sets the yields smartly so that the detection statistics on Bob's side remains the same as the case without attacks.} \label{Fig:USD:TheSetup}
\end{figure}

Eve's attack is composed of two parts, a USD measurement and a PNS attack strategy \cite{BLMS:PNS:2000}, as shown in Fig.~\ref{Fig:USD:TheSetup}. First, Eve performs a positive operator-valued measurement (POVM) to distinguish between a signal state and a decoy state without disturbing the quantum state sent by Alice (see Appendix \ref{AppSection:USDmeasurement} for details). Then she measures the photon number. Conditioned on her measurement results for the signal/decoy information, Eve may forward some photons to Bob so as to preserve the statistics of a normal quantum channel. For a single-photon state, Eve either blocks it or directly forwards it without knowing the qubit information, while for a multiphoton state, Eve can keep one copy and forward the rest giving her the full qubit information (see Appendix \ref{AppSection:PNSattack} for details).
%blocked or partially forwarded with Eve holding one copy of the qubit information.

%\section{Results}

%{\bf Theoretical basis}
An attack is considered successful if Eve is able to trick Alice and Bob into accepting an insecure key. To show that our attack is successful, we compare the following two key rates: (i) a lower bound on the secure key rate from the perspective of Alice and Bob, who overlook phase randomization and apply the conventional decoy-state postprocessing, denoted $R^l$, and (ii) an upper bound on the secure key rate taking into account our attack, denoted $R^u$. The former situation assumes that phase randomization is performed (but is actually not performed) and uses the post-processing scheme presented in Ref.~\cite{MXF:Practical:2005}; the key rate lower bound for this case (details given in Appendix \ref{AppSection:Onedecoy}) is
\begin{equation} \label{USD:Hack:GLLP}
R^l= -Q_{\mu}H(E_{\mu})+Y_{1}^s\mu e^{-\mu}[1-H(e_1^s)], \\
\end{equation}
where $Q_{\mu}$ and $E_{\mu}$ are the overall gain and quantum bit error rate (QBER); $\mu$ denotes the expected photon number of the signal state; $Y_1^s$ and $e_1^s$ are the yield and the error rate of the single-photon signal state, respectively, which are estimated by the decoy-state method; and $H(e)=-e\log_{2}(e)-(1-e)\log_{2}(1-e)$ is the binary Shannon entropy function. Here, we assume that Alice and Bob run the efficient BB84 \cite{Lo:EffBB84:2005} and take the basis sift factor to be 1. On the other hand,
our USD+PNS attack sets an upper bound on the key rate
%by considering our USD+PNS attack, we derive the upper bound of key rate
(details shown in Appendix \ref{AppSection:Upperbound}):
\begin{equation} \label{USD:Hack:KeyRate}
R^{u} = Y_1^se^{-\mu}\mu .
\end{equation}
Note that different values of $Y_1^s$ are used in the computation of $R^{l}$ and $R^{u}$. The value of $Y_1^s$ used in the lower bound, $R^{l}$, is the one estimated by Alice and Bob using conventional decoy-state processing and the value used in the upper bound, $R^{u}$, is the one chosen by Eve in the attack. Since the lower bound represents the key rate at which Alice and Bob generate a new key that they think is secure, if this rate is
higher than what is allowed after taking into account our attack, some of the new key must be insecure and Eve has some information about it. Thus, our attack is successful if
\begin{equation} \label{USD:Hack:SuccessRul}
R^l>R^u \hspace{.5cm} \text{(attack successful)}.
\end{equation}
%%If we can show that $R^l>R^u$, the attack is successful.
%In the following, we show that this inequality indeed holds %using the experimental values obtained
%in our experiment.

Eve's attack aims to preserve the measurement statistics of Bob in order to conceal the attack. We form the attack strategy as an optimization problem subject to preserving the gain statistics. On the other hand, we do not constrain the error statistics here since the error rate introduced by our attack demonstration is negligible. To get more details of this discussion, please see Appendix \ref{AppSection:ErrStat}.
%(see Table \ref{Tab:App:ExpResult} and note that $\xi_\mu$ and $\xi_\nu$ are very close to one).

\section{Experiment setup}

%Our hacking strategy consists of a USD measurement and a PNS attack.
The PNS attack of our USD+PNS hacking strategy, which requires a quantum non-demolition measurement \cite{Imoto:QND:1985,Brune:QND:1990,Holland:QND:1991,Grangier:QND:1998} on the photon numbers, a lossless channel and the ability to control Bob's detector efficiency, is beyond current technology. Thus, we assume in the analysis that Eve has the ability to perform the PNS part and we only realize the USD measurement in our experiment.

\begin{figure*}[tbh]
\centering \resizebox{16cm}{!}{\includegraphics{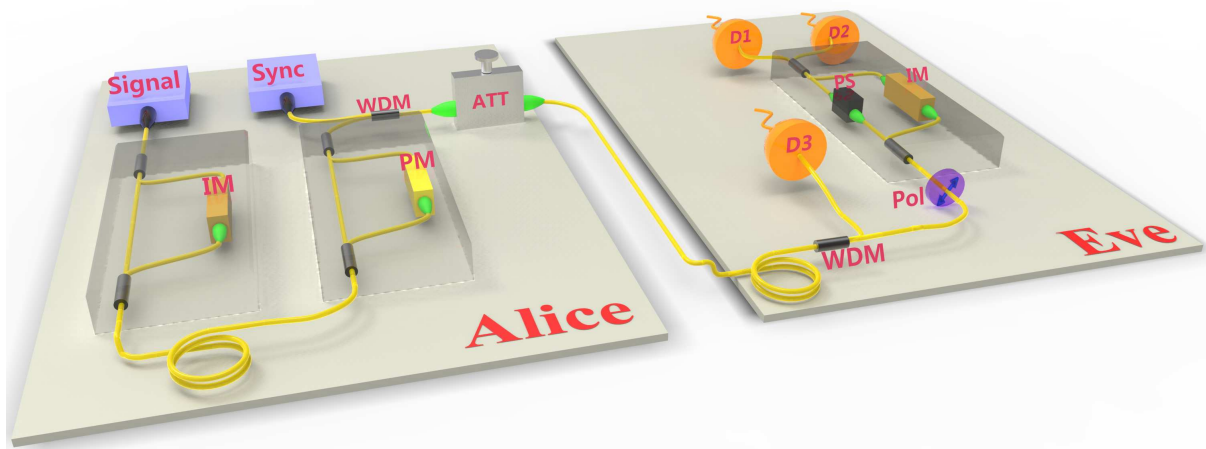}}
\caption{(Color online) Schematic diagram of our experiment setup for USD measurement demonstration. Alice's first AMZI splits the signal pulse into two pulses: a strong one for phase stabilization and a weak one for the quantum signal modulated by the intensity modulator (IM) to be either a decoy state or a signal state. The second AMZI encodes the BB84 states with a phase modulator (PM). Then a synchronization pulse is coupled with the signal pulses into a signal fiber and sent to Bob. At Eve's site, she utilizes a polarization controller and polarizer (Pol) to compensate for the polarization change and a phase shifter (PS) to compensate for the phase drift. Then she uses the same AMZI setup as Alice's first one to make the quantum pulse interfere with the strong phase-stabilization pulse modulated by the IM. The interference result either indicates the identity of the quantum pulse (signal or decoy) or is inconclusive.} \label{Fig:USD:ExpSetup}
\end{figure*}

To demonstrate the attack, we use a phase-encoding BB84 QKD system with strong phase-stabilization pulses \cite{Yuan:Continuous:2005,TYChen:2009:fieldtest,Peev:SECOQC:2009}. The experimental setup is shown in Fig.~\ref{Fig:USD:ExpSetup}. At Alice's site, a distributed feedback (DFB) diode with a central wavelength of 1550.12 nm and a pulse duration of 1 ns operates at a repetition rate of 4 MHz. Then the laser pulse passes through two asymmetric Mach-Zehnder interferometers (AMZIs). The first one splits the coherent pulse generated by the laser source into two time bins, one for the phase stabilization and the other for the quantum signal. Then, the decoy state is prepared by randomly modulating the intensity of the quantum signal. The intensities of the signal and decoy states, $\mu$ and $\nu$, are set to $0.5$ and $0.1$ respectively. In the second AMZI, the phase modulator encodes the quantum pulse with one of the four BB84 phases, and it performs no action on the strong phase-stabilization pulse.

At Eve's site, both the phase-stabilization pulse and the quantum signal are split into two pulses with a 99:1 beam splitter of Eve's AMZI, which have the same splitting ratio and the same path-length difference as Alice's first AMZI. The fraction of the phase-stabilization light, passing through the arm with the intensity modulator, interferes with the fraction of the quantum signal, passing through the arm with the phase shifter. With the intensity modulator, Eve can choose to measure either the signal or the decoy state. %The detailed POVM description is presented in Methods.
%an intensity corresponding to either the signal or decoy state and modulates the strong phase-stabilization pulse with the intensity modulator in the longer arm using this intensity. In effect, each USD measurement attempts to match only one intensity at a time.
%Eve actively chooses one of the two POVMs by modulating the strong
%reference pulse with the IM in the longer arm so that it can choose the
%intensities for the signal and decoy pulse.
Eve's AMZI cancels the path delay between the phase-stabilization pulse and the quantum signal pulse, which is set by Alice's first AMZI, and makes the two pulses meet and interfere with each other.
%The essence of the interferometer is that, since the phase-stabilization pulse precedes the quantum signal pulse with a path delay set by Alice's first AMZI, the same AMZI of Eve cancels the delay and makes the two pulses meet and interfere with each other.
Identical AMZIs will make perfect interference, and in our experiment, we have achieved a high visibility of 500:1.
%{\bf (more explanation here? as suggested by liuyang on p. 6 line -3.)}
The interference results can be divided into two cases: (i) if the pulses in the two arms have the same intensity, detector D2 does not click and the USD measurement result is inconclusive; and (ii) when detector D2 clicks, Eve can identify the state sent by Alice.
%Whether it is a signal or decoy state depends on the anti-correlated relation between the modulation conditions of Alice's and Eve's intensity modulators.
The second case corresponds to a successful USD measurement outcome for Eve.

%See the aforementioned POVM in Eqs.~\eqref{USD:Hack:POVMSignal} and ~\eqref{USD:Hack:POVMDecoy}, and Table \ref{Tab:USD:POVMProbDist}.
\begin{table}
\centering
\caption{Ideal and experimental probabilities of the USD attack, conditioned on signal/decoy states sent by Alice.
% Here, $q_\text{max}$ is defined in Eq.~\eqref{eqn-qmax}.
} \label{Tab:USD:POVMProbDist}
\begin{tabular}{c||ccc|ccc}
\hline
\hline
State of
&
\multicolumn{3}{c}{Ideal case}
&
\multicolumn{3}{|c}{Experimental case}
\\
Alice
& Signal & Decoy & Failure & Signal & Decoy & Failure
%& $\hat{E_{s}^{\mu}}$ & $\hat{E_{f}^{\mu}}$ & $\hat{E_{s}^{\nu}}$ & $\hat{E_{f}^{\nu}}$
\\
\hline
Signal
& $q_\text{max}$ \: & 0 & $1-q_\text{max}$
& $q_\mu\xi_{\mu}$ & $q_\mu(1-\xi_{\mu})$ \:& $1-q_\mu$ \\
%\hline
%decoy & 0 & 1 & $p_{\mu\nu}$ & $1-p_{\mu\nu}$ \\
Decoy & 0 & \: $q_\text{max}$ \: & $1-q_\text{max}$
&
$q_\nu(1-\xi_{\nu})$ & $q_\nu\xi_{\nu}$ & $1-q_\nu$
\\
\hline
\hline
\end{tabular}
\end{table}

\section{Results}

The performance of our USD experiment is characterized by two sets of parameters, as listed in Table \ref{Tab:USD:POVMProbDist}. The first one is related to success probabilities. Since the overlap between the signal and decoy states is nonzero, Eve's USD measurement cannot succeed with unity probability and we denote the success probability when Alice sends a signal (decoy) state $q_{\mu}$ ($q_{\nu}$).
%, where $\mu$ ($\nu$) denotes the expected photon number of the signal (decoy) state.
The second set of parameters is related to error probabilities.
Note that even when the USD measurement succeeds, experimental imperfection may cause the USD to report the wrong state. We denote the probability of correctly identifying the input state conditioned on a successful USD by $\xi_\mu$ ($\xi_\nu$) when Alice sends the signal (decoy) state. These key parameters $q_\mu$, $q_\nu$, $\xi_\mu$, and $\xi_\nu$ characterize the effectiveness of our USD attack from Eve's perspective and her ability to compromise the security of the QKD system. Details of these definitions can be found in Appendix \ref{AppSection:Upperbound}.

Several aspects of our experiment affect the success probabilities $q_{\mu}$ and $q_{\nu}$.
% in our demonstration.
%The success probabilities $q_{\mu}$ and $q_{\nu}$
%in our demonstration
%are limited by the fundamental indistinguishability of non-orthogonal quantum states.
First is the fundamental indistinguishability of nonorthogonal quantum states.
Our USD measurement acts only on the first pulse, and not on the second pulse, which encodes the phase information. The optimal USD to distinguish the two possibilities of the first pulse,
$\ket{\sqrt\frac{\mu}{2}}$ and $\ket{\sqrt\frac{\nu}{2}}$,
has maximal success probability (details given in Appendix \ref{AppSection:USDmeasurement})
\begin{equation}
\label{eqn-q-optimal}
q_\text{opt}=1-
\exp \left( -\frac{1}{4} \left\lvert \sqrt{\nu}-\sqrt{\mu} \right\rvert^2 \right).
%
%\Big\lvert \braket{\sqrt\frac{\nu}{2}}{\sqrt\frac{\mu}{2}} \Big\rvert.
\end{equation}
On the other hand, our linear-optics USD setup, shown in Fig.~\ref{Fig:USD:ExpSetup}, achieves only $q_\text{max}={q_\text{opt}}/{2}$ even when the devices are perfect.
%\begin{equation}
%\label{eqn-q-ExpMaximal}
%\begin{aligned}
%q_\text{max} %&\triangleq \bra{\sqrt{\frac{\mu}{2}}} E_{\mu} \ket{\sqrt{\frac{\mu}{2}}} =\bra{\sqrt{\frac{\nu}{2}}} E_{\nu} \ket{\sqrt{\frac{\nu}{2}}} \\
%%%%&= \frac{1-\left\lvert \braket{0}{\sqrt{\frac{\mu}{4}}-\sqrt{\frac{\nu}{4}}} \right\rvert^2}{2}
%%%%\\
%%%%&=
%%%%\frac{1-\exp\left(-\frac{|\sqrt{\mu}-\sqrt{\nu}|^2}{4}\right)}{2}
%%[1-\exp(-(\sqrt{\mu}-\sqrt{\nu})^2/2)]/2
%%=1.87\%
%%%%\\
%%%%&< q_\text{opt}.
%&=\frac{q_\text{opt}}{2}.
%\end{aligned}
%\end{equation}
It is an interesting question with regard to how to implement a USD measurement to achieve $q_\text{opt}$, especially with linear optics.
When $\mu=0.5$ and $\nu=0.1$, one obtains $q_\text{max}=1.87\%$ and $q_\text{opt}=3.75\%$.
Experimental imperfections and inefficient detectors
%may
further reduce the actual success probability.

The measurement results for $q_{\mu}$, $q_{\nu}$, $\xi_\mu$, and $\xi_\nu$ over a time period of 748 s are listed in Table \ref{Tab:App:ExpResult}. Note that $\xi_\mu$ ($\xi_\nu$) is near 100\%, which indicates that we almost never made a mistake in identifying the state.

\begin{table}
\centering
\caption{Experimental results. The standard deviations (in the time domain) of $q_\mu$, $q_\nu$, $\xi_\mu$, and $\xi_\nu$ are, respectively, $4.3\times10^{-5}$, $4.0\times10^{-5}$, $0.98\%$, and $0.40\%$, which shows that the attack system is robust.} \label{Tab:App:ExpResult}
\begin{tabular}{ccccccc}
\hline
$q_\mu$ & $q_\nu$ & $\xi_\mu$ & $\xi_\nu$ & $q_\text{opt}$ \\
\hline
$1.18\times10^{-3}$ & $1.16\times10^{-3}$ & $96.90\%$ & $98.37\%$ &
$3.75\%$
%$7.08435\%$
%$7.35\%$
\\
\hline
\end{tabular}
\end{table}

\begin{figure}[tbh]
\centering \resizebox{12cm}{!}{\includegraphics{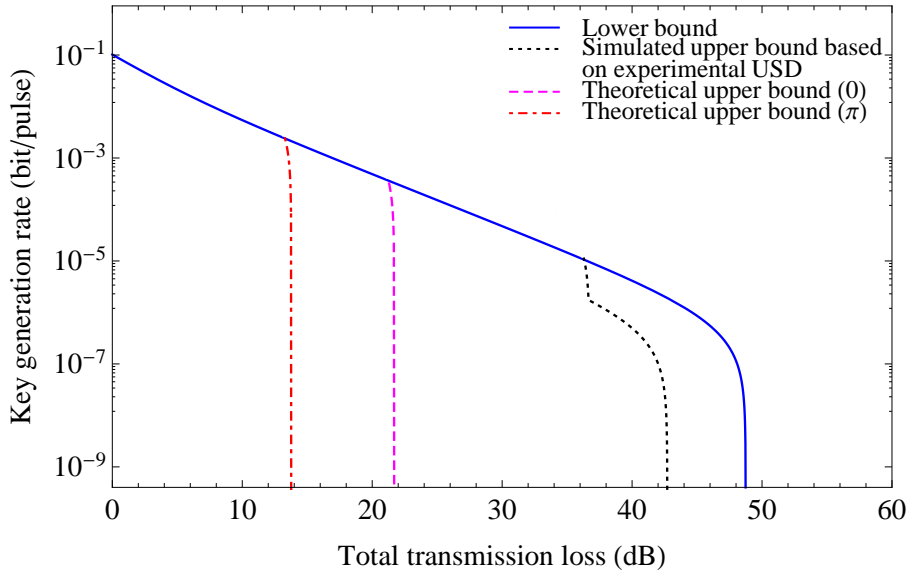}}
\caption{(Color online) Bounds on the key generation rate. The lower bound, given in Eq.~\eqref{USD:Hack:GLLP}, is computed by ignoring the phase randomization problem. The simulated upper bound is evaluated by Eq.~\eqref{USD:Hack:KeyRate} and the data listed in Table \ref{Tab:App:ExpResult}. The region for which the upper bound is below the lower bound corresponds to the secret key being stolen by Eve. Also shown are two best theoretical upper bounds using ideal values $q_\mu=q_\nu=23.0\%,3.75\%$ and $\xi_\mu=\xi_\nu=1$, where the dash-dotted (dashed) curve corresponds to a setup with a relative phase of $\pi$ ($0$) between signal and decoy pulses giving rise to $q_\mu=q_\nu=23.0\%$ ($3.75\%$). Upper bounds corresponding to other relative phases fall between these two curves. Note that in our experiment, the relative phase is 0. Our attack is successful when the lower bound is higher than the upper bound, which occurs when the transmission loss is larger than $36.3$ dB (for our experiment),
$21.2$ dB (for the ideal situation with zero relative phase), and $13.3$ dB (for the ideal situation with $\pi$ relative phase).}
\label{Fig:key-rate-bounds-experimental}
\end{figure}

Using the experimental values for these parameters, we can derive the key rate upper bound as a function of the transmission loss between Alice and Bob, which is shown in Fig.~\ref{Fig:key-rate-bounds-experimental}. Also shown is the the lower bound of the key rate that Alice and Bob thought to be achievable with the assumption of phase randomization, which adopts some realistic parameters of Bob's setup with superconducting single-photon detectors \cite{Yoshino:2012:QKD}: dark count $Y_0=10^{-7}$, detection efficiency $5\%$, and misalignment error rate $e_d=2.0\%$. The key result is that when the overall transmission loss is beyond $36.3$ dB, the upper bound is below the lower bound as shown in Eq.~\eqref{USD:Hack:SuccessRul}, and thus our attack allows Eve to successfully steal the secret key. %We remark that the measured values of $q_{\mu}$, $q_{\nu}$, $\xi_\mu$, and $\xi_\nu$ are valid for all transmission distances since they are obtained by immediately measuring the pulses that have just left Alice's site.

%We show that by exploiting the phase information of the signal and decoy states, our experimental attack succeeds in stealing the final secret key when the transmission loss is over a certain threshold ($36.3$ dB). This is proved by showing a violation in the normal condition that the lower bound on the key rate should be lower than the upper bound. We prove that phase randomization cannot be neglected in decoy-state QKD using WCS, unless a new security proof is available.

To illustrate the potential power of the ideal USD+PNS attack, we consider the ideal USD measurement and take the success probabilities $q_{\mu}$ and $q_{\nu}$ to be the theoretically maximum of $3.75\%$ and the correct distinguishing probabilities, $\xi_{\mu}$ and $\xi_{\nu}$, to be 1. This upper bound curve is shown in Fig.~\ref{Fig:key-rate-bounds-experimental}, which indicates that when the overall loss between Alice and Bob is only beyond $21.2$ dB, the decoy-state BB84 protocol with phase-nonrandomized WCS is insecure. For comparison, an upper bound curve for the success probability of $23.0\%$ corresponding to the relative phase between signal and decoy pulses of $\pi$ is also shown. In essence, this figure shows that the potential impact of our attack can be significant and phase randomization cannot be neglected in decoy-state QKD.

\section{Discussion and conclusion}

By exploiting the phase information of the signal and decoy states, our experimental attack succeeds in stealing the final secret key when the transmission loss is over a certain threshold. We prove that phase randomization cannot be neglected in decoy-state QKD using WCS, unless a new security proof is available. Our result also answers a long-standing question. Before our work, it was unclear whether performing phase randomization improves the key rate performance of decoy-state BB84 using a WCS. Our result implies that performing phase randomization is strictly better.
We remark that our attack is not limited to the phase-encoding system with strong phase-stabilization pulses \cite{Yuan:Continuous:2005,TYChen:2009:fieldtest,Peev:SECOQC:2009} on which our experiment is based. As long as the phase of each state, be it a signal or a decoy state, is known by Eve, she does not need the strong phase reference from Alice. Eve can simply prepare an auxiliary pulse with the corresponding phase. Therefore, this attack can be launched to hack a regular decoy-state QKD system without phase randomization.

%Our attack aims to preserve the measurement statistics of Bob in order to conceal the attack. We form our attack strategy as an optimization problem subject to preserving the gain statistics (see Section 1 of Supplementary Material). On the other hand, we do not constrain the error statistics since the error introduced by our attack comes from the error made experimentally in the USD measurements and this error probability is negligible because $\xi_\mu$ and $\xi_\nu$ are very close to one (see Table \ref{Tab:App:ExpResult}).

A key feature of our experiment is the implementation of USD with linear optics.
%Even though it is based on linear optics and is easy to implement,
Even with only linear optics, this attack system can efficiently compromise the security of the key.
Our work on applying USD with linear optics in quantum information opens an avenue to full linear-optics-based implementation of general quantum measurements, extending previous results \cite{Clarke:USD:2001,vanEnk:USD:2002,Wittmann:USD:2010}.
%Our work is of applying USD with linear optics in quantum information opens an avenue to fully linear-optics-based implementation of general quantum measurements as a powerful technique, extending the previous results \cite{Clarke:USD:2001,vanEnk:USD:2002,Wittmann:USD:2010}.
%Several aspects of our USD experiment renders the USD success probability suboptimal. Firstly, our proof-of-concept USD experiment is not an implementation of the optimal USD measurement (see Section I of Supplemental Material).Secondly, losses in Eve's polarization controller, AMZI and the detector further reduce the intensity and hence the success probability.
For future work, it is an interesting perspective topic to study the case where Eve knows partial information on the phases. For example, in a QKD system with active phase randomization \cite{Zhao:Active:2007}, the phase may not be perfectly random in practice. A related question will be whether a fully randomized phase is necessary for Alice and Bob to guarantee the security.

%[mxf: delete the following para]
%
%{\bf (This paragraph seems difficult to explain clearly, esp the 2nd point.  Delete?)}
%There are some remaining aspects that should be paid attention to. 1)The first case might be that the phase is partial randomized. In this case, the maybe a partial USD measurement should be taken. {\bf (What is partial USD?)} 2)The second case is that, in practice, even when Eve obtains a successful measurement, she might deduce the signal or decoy state wrong. This case followed with PNS attack might probably introduce influence upon phase-encoding or polarization-encoding scheme differently.
%

\section{Acknowledgments}
We thank Y.~Cao,  W.~F.~Cao, L.~J.~Wang, and F.~Zhou for enlightening discussions. This work was supported by National Natural Science Foundation of China, the National Basic Research Program of China Grants, the CAS and USTC Special Grant for Postgraduate Research, Innovation and Practice, and Quantum Communication Technology Co., Ltd., Anhui. X.~M.~gratefully acknowledges the financial support from the National Basic Research Program of China Grants No.~2011CBA00300 and No.~2011CBA00301, National Natural Science Foundation of China Grant No.~61033001; and the 1000 Youth Fellowship program in China. C.-H.~F.~F.~gratefully acknowledges the financial support of RGC Grant No.~700712P from the HKSAR Government.

\begin{appendix}

\section{USD measurement} \label{AppSection:USDmeasurement}
Considering the one-decoy state protocol \cite{MXF:Practical:2005}, where two coherent states are used by Alice, a signal state, $\ket{\sqrt{\mu}e^{i\theta_s}}$, and a weak decoy state, $\ket{\sqrt{\nu}e^{i\theta_d}}$. Here, $\mu$ and $\nu$ are the intensities of the signal and decoy states, respectively, with $\mu>\nu$; $\theta_s$ and $\theta_d$ are the phases of the signal and decoy states, respectively. Since no phase randomization is performed and Alice does not intentionally put any phase difference between signal and decoy pulses, their phases are naturally the same and thus we take $\theta_s=\theta_d=0$.
%\footnote{For the case where $\theta_s-\theta_d\neq0$, the success probability for Eve's attack will increase as shown in Fig.~\ref{Fig:key-rate-bounds-experimental}.}.
(For the case where $\theta_s-\theta_d\neq0$, the success probability for Eve's attack will increase as shown in Fig.~\ref{Fig:key-rate-bounds-experimental}.)
%~\ref{Fig:key-rate-bounds-experimental}).

For the phase-encoding scheme, Alice encodes qubits in the relative phases of the two pulses separated by the second Mach-Zehnder interferometer, as shown in Fig.~\ref{Fig:USD:ExpSetup}.
%~\ref{Fig:USD:ExpSetup}.
For the BB84 protocol, she encodes $\phi\in\{0,\pi/2,\pi,3\pi/2\}$ randomly and sends out,
\begin{equation}
\label{USD:Hack:States}
\begin{aligned}
\ket{\psi_s} &= \ket{\sqrt\frac{\mu}{2}}\ket{e^{i\phi}\sqrt\frac{\mu}{2}}, \\
\ket{\psi_d} &= \ket{\sqrt\frac{\nu}{2}}\ket{e^{i\phi}\sqrt\frac{\nu}{2}}, \\
\end{aligned}
\end{equation}
for signal and decoy states, respectively. %For each state sent by Alice, a random phase $\phi$ is chosen.

Since the signal and decoy states are not orthogonal (i.e., $\braket{\psi_s}{\psi_d}\neq0$), Eve cannot perfectly distinguish them with unity probability. Instead, she performs a USD measurement to perfectly distinguish them with probability less than one.
We impose that our USD measurement acts only on the first pulse and we leave the second pulse, which encodes the phase information, intact. This is because a measurement on the second pulse may destroy the qubit encoded in the relative phase.
%a USD measurement on both pulses is experimentally challenging.
In this case, the failure probability corresponding to the optimal USD \cite{Ivanovic:1987:USD,Dieks:1988:USD,Peres:1988:USD} (assuming equal \emph{a prior} probabilities of $\ket{\psi_s}$ and $\ket{\psi_d}$, which is the case in our experiment) is given by the overlap between the states to be identified:
\begin{equation} \label{USD:Hack:FailProb}
\begin{aligned}
p_f &=
\Big\lvert \braket{\sqrt\frac{\nu}{2}}{\sqrt\frac{\mu}{2}} \Big\rvert
%\sum_{\phi=0,\pi/2,\pi,3\pi/2}\braket{\psi_s}{\psi_d}
\\
&=
\exp \left( -\frac{1}{2} \left\lvert \sqrt\frac{\nu}{2}-\sqrt\frac{\mu}{2} \right\rvert^2 \right).
\end{aligned}
\end{equation}
The corresponding POVM is
\begin{equation} \label{USD:Hack:POVM}
\begin{aligned}
\hat{E_{\mu}}&=
\frac{1}{(1+p_f)(1-p_f^2)}
P(\ket{\sqrt{\frac{\mu}{2}}} - \braket{\sqrt{\frac{\nu}{2}}}{\sqrt{\frac{\mu}{2}}} \ket{\sqrt{\frac{\nu}{2}}}), \\
%P(\ket{\sqrt{\mu}} - \braket{\sqrt{\nu}}{\sqrt{\mu}} \ket{\sqrt{\nu}}) \\
%\frac{I-\ket{\sqrt{\nu}}\bra{\sqrt{\nu}}}
%{1+\lvert \braket{\sqrt{\nu}}{\sqrt{\mu}} \rvert} \\
\hat{E_{\nu}}&=
\frac{1}{(1+p_f)(1-p_f^2)}
P(\ket{\sqrt{\frac{\nu}{2}}} - \braket{\sqrt{\frac{\mu}{2}}}{\sqrt{\frac{\nu}{2}}} \ket{\sqrt{\frac{\mu}{2}}}),
%P(\ket{\sqrt{\nu}} - \braket{\sqrt{\mu}}{\sqrt{\nu}} \ket{\sqrt{\mu}})
%\hat{E_{\nu}} &=\frac{1}{1+p_f}
%(I-\ket{\sqrt{\mu}}\bra{\sqrt{\mu}})
\\
\hat{E_{f}}&=
I-\hat{E_{\mu}}-\hat{E_{\nu}},
%\frac{\ket{\sqrt{\nu}}\bra{\sqrt{\nu}}}{2}, \\
\end{aligned}
\end{equation}
where
we define the projection function $P(\ket{\varphi})=\ket{\varphi}\bra{\varphi}$ for some state $\ket{\varphi}$.
% and $p_f=\lvert \braket{\sqrt{\nu}}{\sqrt{\mu}} \rvert$ is the failure probability.
The measurement outcome $\hat{E_{\alpha}}$ indicates that the input state is $\ket{\sqrt{\alpha/2}}$, where $\alpha=\mu,\nu$; and the outcome $\hat{E_{f}}$ is inconclusive about whether the input state is a signal state or a decoy state.
Note that the success probability corresponding to the optimal USD is
\begin{equation}
\label{eqn-qmax}
q_\text{opt}\triangleq\bra{\sqrt\frac{\mu}{2}} \hat{E_{\mu}} \ket{\sqrt\frac{\mu}{2}} =\bra{\sqrt\frac{\nu}{2}} \hat{E_{\nu}} \ket{\sqrt\frac{\nu}{2}}=
%1-
%\Big\lvert \braket{\sqrt\frac{\nu}{2}}{\sqrt\frac{\mu}{2}} \Big\rvert
%\lvert \braket{\sqrt{\nu}}{\sqrt{\mu}} \rvert,
1-p_f,
\end{equation}
and error does not occur since $\bra{\sqrt{\mu/2}} \hat{E_{\nu}} \ket{\sqrt{\mu/2}} =\bra{\sqrt{\nu/2}} \hat{E_{\mu}} \ket{\sqrt{\nu/2}}=0$.
The ideal probabilities of this optimal USD measurement outcomes
%these POVM operators
are summarized in Table \ref{Tab:USD:POVMProbDist}.
%\ref{Tab:USD:POVMProbDist}.
%For instance,
%%when the POVM of Eq.~\eqref{USD:Hack:POVMSignal}
%when the USD measurement for the signal state corresponding to
%Eq.~\eqref{USD:Hack:POVMSignal}
%is used, the measurement outcome $\hat{E_{s}^{\mu}}$ tells  Eve that the state is a signal state, $\ket{\sqrt{\mu}}$, and
%the measurement outcome $\hat{E_{f}^{\mu}}$ indicates an inconclusive result that the state could either be a signal or decoy state.
In our experiment, we implement the USD measurement with linear optics as shown in Fig.~\ref{Fig:USD:ExpSetup}, and its maximal success probability, assuming 100\% efficiency detectors, is given by
%$q_\text{max}$
\begin{equation}
\label{eqn-q-maximal}
\begin{aligned}
q_\text{max} %&\triangleq \bra{\sqrt{\frac{\mu}{2}}} E_{\mu} \ket{\sqrt{\frac{\mu}{2}}} =\bra{\sqrt{\frac{\nu}{2}}} E_{\nu} \ket{\sqrt{\frac{\nu}{2}}} \\
&= \frac{1-\left\lvert \braket{0}{\sqrt{\frac{\mu}{4}}-\sqrt{\frac{\nu}{4}}} \right\rvert^2}{2}
\\
&=
\frac{1-\exp\left(-\frac{|\sqrt{\mu}-\sqrt{\nu}|^2}{4}\right)}{2}
%[1-\exp(-(\sqrt{\mu}-\sqrt{\nu})^2/2)]/2
%=1.87\%
\\
%&\leq q_\text{opt}.
&=\frac{q_\text{opt}}{2},
\end{aligned}
\end{equation}
%which is Eq.~(5) in the Letter.
%, given in Eq.~5 in the Letter, is lower than $q_\text{opt}$.
It is an interesting question
%for us
to find a way to implement the optimal USD measurement corresponding to Eq.~\eqref{USD:Hack:POVM} using linear optics.

\section{PNS attack} \label{AppSection:PNSattack}
After the USD measurement, Eve measures the photon number of Alice's pulse and launches the PNS attack. The photon numbers of the two WCSs follow the Poisson distribution:
\begin{equation} \label{USD:Analysis:Poisson}
\begin{aligned}
P_{i}^s = \frac{\mu^ie^{-\mu}}{i!}, \\
P_{i}^d = \frac{\nu^ie^{-\nu}}{i!}. \\
\end{aligned}
\end{equation}
Define the gains, $Q_{\mu}$ and $Q_{\nu}$, respectively, to be the probabilities for Bob to get a detection event given that Alice sends signal and decoy states,
\begin{equation} \label{USD:App:Gains}
\begin{aligned}
Q_\mu &= Y_0^sP_{0}^s+Y_1^sP_{1}^s+Y_2^sP_{2}^s+\cdots, \\
Q_\nu &= Y_0^dP_{0}^d+Y_1^dP_{1}^d+Y_2^dP_{2}^d+\cdots. \\
\end{aligned}
\end{equation}
where $Y_i$ is the yield of the $i$-photon state or the conditional probability for Bob to get a detection given that Alice sends out an $i$-photon states; the superscripts, $s$ and $d$, denote the signal state and decoy state, respectively.

In the post processing of decoy-state QKD, the yield of the single-photon-state component can be informed from the detection statistics of signal and decoy states on Bob's side. The underlying assumption of the photon number channel model for the security proof of decoy-state QKD \cite{Lo:Decoy:2005} can be described as
\begin{equation} \label{USD:App:YieldSame}
\begin{aligned}
Y_i^s = Y_i^d,
\end{aligned}
\end{equation}
which holds when the phases of signal and decoy states are randomized. In our USD+PNS attack, Eq.~\eqref{USD:App:YieldSame} is violated when Eve is able to distinguish between the signal and decoy states by a USD measurement given the phase information of the coherent states. (Note that Eq.~\eqref{USD:App:YieldSame} may also be violated even when only partial phase information is known to Eve \cite{Sun:Partially:2012}.)
She smartly chooses these proportions ($Y_i^s$ and $Y_i^d$) so that her attack will not be detected. This is achieved by maintaining the same observed gain statistics ($Q_\mu$ and $Q_\nu$) as in the normal situation.

%\subsection{Key rate upper bound}
\section{Key rate upper bound}\label{AppSection:Upperbound}
In the USD attack, Eve performs the POVM as shown in Fig.~\ref{Fig:USD:ExpSetup}.
%given in Eq.~\eqref{USD:Experimental-POVM}.
%Eve performs a POVM with four elements, $\{\hat{E_{s}^{\mu}}, \hat{E_{f}^{\mu}}, \hat{E_{s}^{\nu}}, \hat{E_{f}^{\nu}}\}$.
Conditioned on these results, Eve sets a different yield (detection probability) for each $i$-photon state. Define $q_{\mu}$ ($q_{\nu}$) to be the conditional probability for Eve to result in a successful measurement outcome when Alice sends out a signal (decoy) state.
%For simplicity, we assume Alice chooses the signal and decoy states evenly. Then, for Eve's best interest, she would perform measurement $\{\hat{E_{s}^{\mu}}, \hat{E_{f}^{\mu}}\}$ and $\{\hat{E_{s}^{\nu}}, \hat{E_{f}^{\nu}}\}$ evenly as well.
Both experimental success probabilities, $q_{\mu}$ and $q_{\nu}$, are limited by the theoretical maximum,
% overlap between signal and decoy states,
\begin{equation} \label{USD:App:qmax}
\begin{aligned}
q_\mu, q_\nu \le
q_\text{max},
%q_{opt}= \frac{1-|\braket{\sqrt{\nu}}{\sqrt{\mu}}|^2}{2},\\
\end{aligned}
\end{equation}
where
%$q_{opt}$
$q_\text{max}$ given in Eq.~\eqref{eqn-q-maximal}
can be achieved when
%$\theta_s$=$\theta_d$ and
Eve's detection efficiency is 100\%.

In practice, even when Eve obtains a successful measurement outcome, she might make an error in determining whether the state is a signal or a decoy state.
% imperfectly.
Define
\begin{equation} \label{USD:App:probxi}
\begin{aligned}
\xi_\mu &
%\equiv
\triangleq
\text{Prob}(E_{\mu}|signal) \\
\xi_\nu &
%\equiv
\triangleq
\text{Prob}(E_{\nu}|decoy) \\
\end{aligned}
\end{equation}
to be the conditional probabilities for Eve to guess Alice's state correctly when Eve obtains successful measurement outcome and Alice sends out a signal (decoy) state. The relationship between these probabilities when Alice sends out a signal and decoy state is shown in Table~\ref{Tab:USD:POVMProbYieldDist}.
This table appears as part of Table \ref{Tab:USD:POVMProbDist}, which also shows the ideal probabilities of the optimal USD measurement.
%(Note that we include the probability of $1/2$ for choosing each POVM in this table, while such probability is not included in Table \ref{Tab:USD:POVMProbDist}.)

\begin{table}
\centering
\caption{Probabilities of these POVM outcomes, conditioned on different intensity states sent by Alice, and yields for different POVM outcomes and photon numbers $i$.} \label{Tab:USD:POVMProbYieldDist}
\begin{tabular}{c|ccc}
\hline
\hline
POVM & $E_{\mu}$ & $E_{\nu}$ & $E_{f}$ \\
\hline
Signal
%$q_\mu\xi_{\mu}$ & $\frac{1}{2}-q_\mu$ & $q_\mu(1-\xi_{\mu})$ & $\frac{1}{2}$
%\\
& $q_\mu\xi_{\mu}$ & $q_\mu(1-\xi_{\mu})$ \:& $1-q_\mu$ \\
%\hline
Decoy
%& $q_\nu(1-\xi_{\nu})$ & $\frac{1}{2}$ & $q_\nu\xi_{\nu}$ & $\frac{1}{2}-q_\nu$
&
$q_\nu(1-\xi_{\nu})$ & $q_\nu\xi_{\nu}$ & $1-q_\nu$
\\
%\hline
Yields & $Z_i^{\mu}$ & $Z_i^{\nu}$ & $X_i$  \\
\hline
\hline
\end{tabular}
\end{table}

Define $Z_i^{\mu}$, $Z_i^{\nu}$, and $X_i$ to be the yields of the $i$-photon state, conditioned on Eve getting the measurement outcome $E_{\mu}$, $E_{\nu}$, and $E_{f}$, respectively, as listed in the last row in Table \ref{Tab:USD:POVMProbYieldDist}. The yield is the probability that a valid detection occurs when Eve sends a pulse to Bob after the attack.
We assume that Eve sets $Z_0^{\mu}$, $Z_0^{\nu}$, and $X_0$ to 0. This is because if Eve forwards any photon to Bob when she gets a vacuum state, she may introduce errors.
The quantum no-cloning theorem tells us that when the photon number is 1, Eve is unable to keep a copy of the qubit information, then the yields $Z_i^{\mu}$, $Z_i^{\nu}$, and $X_i$ will enable Bob to generate secure keys from this one-photon state component.

In our attack, we set $X_i=0$ for all $i$, for the following reason. Since our implementation of the USD measurement destroys the quantum reference pulse, if the USD outcome is inconclusive (i.e., $E_{f}$), Eve cannot always choose the right intensity for the regenerated reference pulse. This increases the error rate for Bob, which alerts Alice and Bob to Eve's presence. Thus, in order to avoid this, we design Eve's attack so that when she fails to learn the state intensity, she does not forward any pulse to Bob in order to emulate a channel loss. This means that $X_i=0$ for all $i$, which we assume for the remaining analysis. Note that if a nondemolition method is used to identify the intensity, no additional errors are introduced and thus there is no need to set
$X_i^{\mu}=X_i^{\nu}=0$ for all $i$.

Thus, the yields $Y_i^{s}$ and $Y_i^{d}$, from Bob's point of view, are composed of
the two successful outcomes of Eve as listed in Table \ref{Tab:USD:POVMProbYieldDist}:
\begin{equation} \label{USD:Analysis:Yield}
\begin{aligned}
Y_i^s &= q_\mu\left[\xi_\mu Z_i^{\mu} +(1-\xi_\mu)Z_i^{\nu}\right],
\\
%+ \frac{1}{2}X_i^{\mu} + \left(\frac{1}{2}-q_\mu\right)X_i^{\nu}\\
Y_i^d &= q_\nu\left[\xi_\nu Z_i^{\nu} +(1-\xi_\nu)Z_i^{\mu}\right]
%+ \frac{1}{2}X_i^{\nu} + \left(\frac{1}{2}-q_\nu\right)X_i^{\mu}
. \\
\end{aligned}
\end{equation}
%Eq.~\eqref{USD:Analysis:Yield} can be understood as the
Then, by inserting Eq.~\eqref{USD:Analysis:Yield} into Eq.~\eqref{USD:App:Gains}, the gains of signal and decoy states are given by
\begin{equation} \label{USD:App:StatisMaintain}
\begin{aligned}
Q_\mu &= \sum_{i=1}^{\infty}  q_\mu\left[\xi_\mu Z_i^{\mu} +(1-\xi_\mu)Z_i^{\nu}\right]
%+ \frac{1}{2}X_i^{\mu} + \left(\frac{1}{2}-q_\mu\right)X_i^{\nu} \right\}
e^{-\mu}\frac{\mu^{i}}{i!},\\
Q_\nu &= \sum_{i=1}^{\infty} q_\nu\left[\xi_\nu Z_i^{\nu} +(1-\xi_\nu)Z_i^{\mu}\right]
%+ \frac{1}{2}X_i^{\nu} + \left(\frac{1}{2}-q_\nu\right)X_i^{\mu} \right\}
e^{-\nu}\frac{\nu^{i}}{i!} .
\end{aligned}
\end{equation}
For a normal quantum channel, Alice and Bob should get,
\begin{equation} \label{USD:App:GainsSim}
\begin{aligned}
Q_\mu &= 1-e^{-\eta\mu}, \\
Q_\nu &= 1-e^{-\eta\nu}. \\
\end{aligned}
\end{equation}
If Eve does not want to disturb the detection statistics on Bob's side, she should choose $Z_i^{\mu}$
and
%$X_i^{\nu}$,
$Z_i^{\nu}$
%, and $X_i^{\mu}$
smartly, so that Eqs.~\eqref{USD:App:StatisMaintain} and \eqref{USD:App:GainsSim} are satisfied.

In the decoy-state postprocessing \cite{MXF:Practical:2005}, the secure key is only derived from the single-photon component. Then, the upper bound of the key rate is given by \cite{MXF:2way:2006}
\begin{equation} \label{USD:Appendix:KeyRate}
\begin{aligned}
R^{u} &= Y_1^se^{-\mu}\mu \\
&=
%\left\{
q_\mu\left[\xi_\mu Z_1^{\mu} +(1-\xi_\mu)Z_1^{\nu}\right]
%+ \frac{1}{2}X_i^{\mu} + \left(\frac{1}{2}-q_\mu\right)X_i^{\nu} \right\}
e^{-\mu}\mu. \\
\end{aligned}
\end{equation}
%\begin{equation} \label{USD:Appendix:KeyRate}
%\begin{aligned}
%R^{u} &= Y_1^se^{-\mu}\mu \\
%&= \left\{q_\mu\left[\xi_\mu Z_i^{\mu} +(1-\xi_\mu)Z_i^{\nu}\right] + \frac{1}{2}X_i^{\mu} + \left(\frac{1}{2}-q_\mu\right)X_i^{\nu} \right\}e^{-\mu}\mu. \\
%\end{aligned}
%\end{equation}
%%(I'm wondering that, since adopting which intensity to generate the final key by Alice and Bob is after Eve's attack which Eve has no pre-information about, then Eve should ensure that the larger one of the key rates with both the intensities should be taken into account, while considering only one of it is not sufficient.)

Now, Eve needs to optimize $Z_i^{\mu}$ and
%, $X_i^{\nu}$,
$Z_i^{\nu}$
%, and $X_i^{\mu}$
in order to minimize the upper bound of the key rate, Eq.~\eqref{USD:Appendix:KeyRate}. The optimization problem can be stated as follows:
\begin{equation} \label{USD:App:minProblem}
\begin{aligned}
%\min_{\{Z_i^{\mu}, X_i^{\nu}, Z_i^{\nu}, X_i^{\mu}\}} Y_1^s \\
\min_{\{Z_i^{\mu}, Z_i^{\nu}\}} Y_1^s
\end{aligned}
\end{equation}
subject to
\begin{equation} \label{USD:Appendix:GainConstraint}
\begin{aligned}
Q_\mu &= 1-e^{-\eta\mu} =
\sum_{i=1}^{\infty}  q_\mu\left[\xi_\mu Z_i^{\mu} +(1-\xi_\mu)Z_i^{\nu}\right]
e^{-\mu}\frac{\mu^{i}}{i!}\\
%\sum_{i=0}^{\infty} \left\{ q_\mu\left[\xi_\mu Z_i^{\mu} +(1-\xi_\mu)Z_i^{\nu}\right] + \frac{1}{2}X_i^{\mu} + \left(\frac{1}{2}-q_\mu\right)X_i^{\nu} \right\}e^{-\mu}\frac{\mu^{i}}{i!} \\
Q_\nu &= 1-e^{-\eta\nu} =
\sum_{i=1}^{\infty} q_\nu\left[\xi_\nu Z_i^{\nu} +(1-\xi_\nu)Z_i^{\mu}\right]
e^{-\nu}\frac{\nu^{i}}{i!} ,
%\sum_{i=0}^{\infty} \left\{ q_\nu\left[\xi_\nu Z_i^{\nu} +(1-\xi_\nu)Z_i^{\mu}\right] + \frac{1}{2}X_i^{\nu} + \left(\frac{1}{2}-q_\nu\right)X_i^{\mu} \right\}e^{-\nu}\frac{\nu^{i}}{i!}, \\
\end{aligned}
\end{equation}
where all $Z_i^{\mu}$
%, $X_i^{\nu}$,
and $Z_i^{\nu}$
%, and $X_i^{\mu}$
are in the regime $[0,1]$. In the detection statistics equations, $\mu$ and $\nu$ are given by Alice's intensity choice, and $q_\mu$, $q_\nu$ are determined both by the %orthogonal ratio
overlap
between the signal and decoy states and by Eve's USD measurement efficiency.
%Then, for different
For a given overall efficiency $\eta$ between Alice and Bob, we can calculate the key rate upper bound.

\section{Experimental details}\label{AppSection:Expdetails}
In our experimental demonstration, the laser source is produced by a DFB diode with a central wavelength of 1550.12 nm and a pulse duration of 1 ns, operating at a repetition rate of 4 MHz. Alice sets $\mu=0.5$ and $\nu=0.1$, and randomly modulates the signal and decoy states with uniform probabilities. Eve performs the USD measurement for the signal and decoy states randomly with equal probabilities as well.

\begin{figure*}[tbh]
\includegraphics[width=16cm]{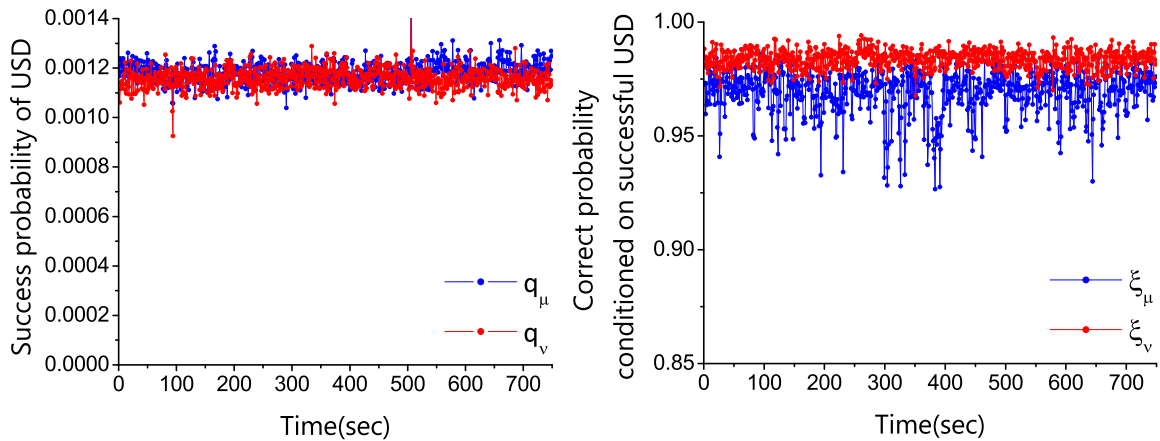}
\caption{(Color online) Experimental results over time for the USD attack demonstration. Note that $\xi_\mu$ ($\xi_\nu$) is close to 100\%, which indicates that we have a nearly perfect interferometer with a high visibility achieving 500:1.}
\label{Fig:USD:ExpResult}
\end{figure*}

The experiment results are collected over an operation time of 748 s. The fluctuation of the attack performance through time is shown in Fig.~\ref{Fig:USD:ExpResult}, where we can see that the results are very stable.

%{\bf The relative phase between signal and decoy states.}
In our experiment, the phases of signal pulses and decoy pulses are both 0 because the phase reference of every pulse is passed to Eve. In some other systems, the relative phase between signal and decoy pulses may not be 0 (i.e., $\theta_s-\theta_d \neq 0$) and it can be shown that the success probability of USD is maximized when the relative phase is $\pi$ and minimized when it is $0$, corresponding to $23.0\%$ and $3.75\%$, respectively.
With a higher success probability, Eve is more capable of stealing the final key and thus the upper bound of the key rate becomes lower. Figure \ref{Fig:key-rate-bounds-experimental}
%\ref{Fig:key-rate-bounds-experimental}
shows the two upper bounds corresponding to the two success probabilities.

\section{Review of one-decoy state} \label{AppSection:Onedecoy}
The key assumption in the security proof of decoy-state QKD \cite{Lo:Decoy:2005} is the equivalence between phase-randomized coherent states and the photon number channel model. A WCS can be described as a superposition of photon number (Fock) states
\begin{equation} \label{USD:App:WCS}
\begin{aligned}
\ket{\alpha} = \ket{\sqrt{\mu}e^{i\theta}} =e^{-|\alpha|^2/2}\sum_{n=0}^\infty \frac{\alpha^n}{\sqrt{n!}} |n\rangle, \\
\end{aligned}
\end{equation}
where $\mu$ and $\theta$ are the intensity and phase of the coherent state, respectively. Since the eavesdropper has no knowledge of phase $\theta$, from her point of view, the density matrix of the state should be written as \cite{Lo:Decoy:2005}
\begin{equation} \label{USD:App:WCS}
\begin{aligned}
\rho_\mu = \int_0^{2\pi} \frac{d\theta}{2\pi} |\sqrt{\mu}e^{i\theta}\rangle\langle
\sqrt{\mu}e^{i\theta}|=e^{-\mu}\sum_{n=0}^\infty  \frac{\mu^n}{n!} |n\rangle\langle n|\\
\end{aligned}
\end{equation}
As shown, the state is a Poisson distributed mixture of photon number state $|n\rangle$.
Then, the channel between Alice and Bob can be understood as a photon number channel. Alice uses channel $n$ with a probability of $e^{-\mu}\frac{\mu^n}{n!}$ to send out an $n$-photon state to carry the qubit information.

Based on the photon number channel model, we briefly review the postprocessing for the one-decoy state protocol \cite{MXF:Practical:2005}. The lower bound of the key rate when Alice and Bob are unaware of
%ignore
Eve's attack, is given by
\begin{equation} \label{USD:App:GLLP}
\begin{aligned}
R^l= -Q_{\mu}H(E_{\mu})+Y_{1}\mu e^{-\mu}[1-H(e_1)], \\
\end{aligned}
\end{equation}
where $Q_{\mu}$ and $E_{\mu}$ are the overall gain and QBER, and $Y_1$ and $e_1$ are the yield and the error rate of the single-photon state, which are estimated by the decoy states. From the analysis using the one-decoy state protocol \cite{MXF:Practical:2005}, one can derive the lower bound of $Y_1$ and upper bound of $e_1$:
\begin{equation}\label{USD:App:OneY1e1Bound}
\begin{aligned}
Y_1 &\ge \frac{\mu}{\mu\nu-\nu^2}(Q_\nu e^{\nu}-Q_\mu
e^\mu\frac{\nu^2}{\mu^2}-E_\mu Q_\mu e^{\mu}\frac{\mu^2-\nu^2}{e_0\mu^2}), \\
e_1 &\le  \frac{E_\mu Q_\mu e^{\mu}}{Y_1^{L,\mu,0}\mu}.
\end{aligned}
\end{equation}

The gains, $Q_\mu$ and $Q_\nu$, are given in Eq.~\eqref{USD:App:GainsSim}. We model the overall QBER in the normal quantum channel to be
\begin{equation} \label{USD:APP:QBER}
E_\mu Q_\mu = e_0 Y_0 + e_d (1 - e^{-\eta \mu}),
\end{equation}
where $e_0=1/2$ is the error rate of the background count; $Y_0$ is the background count rate, which includes the detector dark count and other background contributions; and $e_d$ is the probability that a photon triggers the incorrect detector and is due to the misalignment and instability of the optical system. As we did not implement Bob's system in our experiment, we adopt some realistic parameters of a setup with
superconducting single-photon detectors \cite{Yoshino:2012:QKD}: $Y_0=10^{-7}$, $e_d=2.0\%$, and a detection efficiency of $5\%$.
%up-conversion single-photon detectors \cite{Liu:MIQKDexp:2012}: $Y_0=1 \times 10^{-6}$, $e_d=0.1\%$, and a detection efficiency of $20\%$.

\section{Error statistics Analysis}\label{AppSection:ErrStat}
In this attack, we form the attack strategy as an optimization problem subject to preserving the gain statistics without maintaining the error statistics, since the attack induces only a low error rate and is not noticeable by looking at the error rate statistics. Here we analyze in detail the more rigorous results of the key rate upper bound, when Eve strictly maintains the gain statistics and the error statistics simultaneously. We can see that even considering the error rate introduced by our attack demonstration that Alice and Bob might check strictly, Eve can still successfully steal the secure key in the same channel loss regime, only with the key amount that Eve can steal compromised.

To maintain the error statistics, Eve should satisfy the equations:
\begin{equation} \label{USD:Appendix:ErrorConstraint}
\begin{aligned}
E_\mu Q_\mu &= \frac{1}{2} Y_0 + e_d(1-e^{-\eta\mu}) \\
&=\frac{1}{2} Z_0^\mu + \sum_{i=1}^{\infty}  q_\mu\left[\epsilon_i^\mu \xi_\mu Z_i^{\mu} + \frac{1}{2} (1-\xi_\mu)Z_i^{\nu}\right]
e^{-\mu}\frac{\mu^{i}}{i!}\\
%\sum_{i=0}^{\infty} \left\{ q_\mu\left[\xi_\mu Z_i^{\mu} +(1-\xi_\mu)Z_i^{\nu}\right] + \frac{1}{2}X_i^{\mu} + \left(\frac{1}{2}-q_\mu\right)X_i^{\nu} \right\}e^{-\mu}\frac{\mu^{i}}{i!} \\
E_\nu Q_\nu &= \frac{1}{2} Y_0 + e_d(1-e^{-\eta\nu})\\
&=\frac{1}{2} Z_0^\nu + \sum_{i=1}^{\infty} q_\nu\left[\epsilon_i^\nu \xi_\nu Z_i^{\nu} + \frac{1}{2} (1-\xi_\nu)Z_i^{\mu}\right]
e^{-\nu}\frac{\nu^{i}}{i!} ,
%\sum_{i=0}^{\infty} \left\{ q_\nu\left[\xi_\nu Z_i^{\nu} +(1-\xi_\nu)Z_i^{\mu}\right] + \frac{1}{2}X_i^{\nu} + \left(\frac{1}{2}-q_\nu\right)X_i^{\mu} \right\}e^{-\nu}\frac{\nu^{i}}{i!}, \\
\end{aligned}
\end{equation}
where $\epsilon_i^{\mu(\nu)}$ is the error Eve sets when Alice sends a signal (decoy) state and Eve gets the correct USD measurement result, and $Z_0^{\mu(\nu)}$ is the dark count Eve sets when there is no photon in the signal (decoy) state Alice sends. Note that Eve simply sets the QBER error to be the upper bound $\frac{1}{2}$ here, when Eve gets the incorrect USD measurement results.

\begin{figure}[tbh]
\centering \resizebox{12cm}{!}{\includegraphics{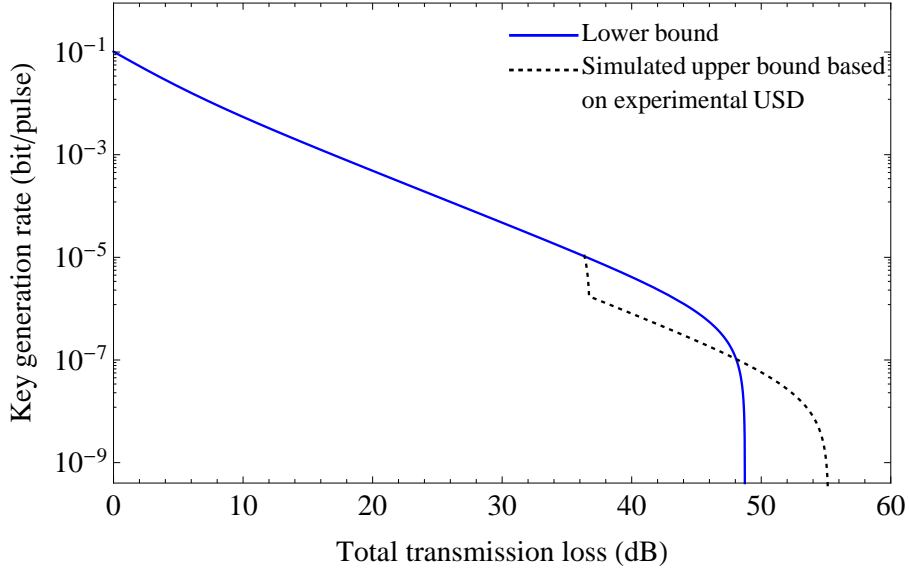}}
\caption{(Color online) Bounds on the key generation rate.
%The lower bound
%, given in Eq.~\eqref{USD:Hack:GLLP},
%is computed by ignoring the phase randomization problem. The simulated upper bound is evaluated by Eq.~\eqref{USD:Hack:KeyRate} and the data given in Table \ref{Tab:App:ExpResult}. The region for which the upper bound is below the lower bound corresponds to the secret key being stolen by Eve. Also shown are two best theoretical upper bounds using ideal values $q_\mu=q_\nu=23.0\%,3.75\%$ and $\xi_\mu=\xi_\nu=1$, where the dash-dotted (dash) curve corresponds to a setup with a relative phase of $\pi$ ($0$) between signal and decoy pulses giving rise to $q_\mu=q_\nu=23.0\%$ ($3.75\%$). Upper bounds corresponding to other relative phases fall between these two curves. Note that in our experiment, the relative phase is zero.
Our attack is successful when the lower bound is higher than the upper bound, which occurs when the transmission loss is between $36.3$ and $48.1$ dB (for our experiment),
%$21.2$ dB (for the ideal situation with zero relative phase), and $13.3$ dB (for the ideal situation with $\pi$ relative phase).
}
\label{Fig:bounds3_ErrorStat3}
\end{figure}

Similarly to Appendix \ref{AppSection:Upperbound}, the optimization problem of minimizing the key rate upper bound can be stated as follows:
\begin{equation}
\begin{aligned}
%\min_{\{Z_i^{\mu}, X_i^{\nu}, Z_i^{\nu}, X_i^{\mu}\}} Y_1^s \\
\min_{\{Z_i^{\mu}, Z_i^{\nu}\}} Y_1^s
\end{aligned}
\end{equation}
subject to
\begin{equation} \label{USD:Appendix:GainErrorConstraint}
\begin{aligned}
Q_\mu &= 1-e^{-\eta\mu} =
\sum_{i=1}^{\infty}  q_\mu\left[\xi_\mu Z_i^{\mu} +(1-\xi_\mu)Z_i^{\nu}\right]
e^{-\mu}\frac{\mu^{i}}{i!},\\
%\sum_{i=0}^{\infty} \left\{ q_\mu\left[\xi_\mu Z_i^{\mu} +(1-\xi_\mu)Z_i^{\nu}\right] + \frac{1}{2}X_i^{\mu} + \left(\frac{1}{2}-q_\mu\right)X_i^{\nu} \right\}e^{-\mu}\frac{\mu^{i}}{i!} \\
Q_\nu &= 1-e^{-\eta\nu} =
\sum_{i=1}^{\infty} q_\nu\left[\xi_\nu Z_i^{\nu} +(1-\xi_\nu)Z_i^{\mu}\right]
e^{-\nu}\frac{\nu^{i}}{i!}, \\
E_\mu Q_\mu &= \frac{1}{2} Y_0 + e_d(1-e^{-\eta\mu}) \geq
\sum_{i=1}^{\infty} \frac{1}{2} q_\mu (1-\xi_\mu)Z_i^{\nu} e^{-\mu}\frac{\mu^{i}}{i!},\\
E_\nu Q_\nu &= \frac{1}{2} Y_0 + e_d(1-e^{-\eta\nu}) \geq
\sum_{i=1}^{\infty} \frac{1}{2} q_\nu (1-\xi_\nu)Z_i^{\mu} e^{-\nu}\frac{\nu^{i}}{i!} ,
\end{aligned}
\end{equation}
where all $Z_i^{\mu}$
%, $X_i^{\nu}$,
and $Z_i^{\nu}$
%, and $X_i^{\mu}$
are in the regime $[0,1]$. In the gain and error statistics equations, $\xi_\mu$, $\xi_\nu$ and $q_\mu$, $q_\nu$ are set as the experimental results.
For a given overall efficiency $\eta$ between Alice and Bob, we can calculate the key rate upper bound, as shown in Fig.~\ref{Fig:bounds3_ErrorStat3}.
%(need to add. I do not know how to add the legend in Mathematica...).

Theoretically, since the error rate of USD measurement is 0, the error statistics is easy to maintain or even optimize. Therefore, the gap between the lower bound in Eq.~\eqref{USD:App:GLLP} and the upper bound in Eq.~\eqref{USD:Appendix:KeyRate} will be the same as the one shown in Fig.~\ref{Fig:key-rate-bounds-experimental}.

\end{appendix}

%%%%%%%%%%%%%%%%%%%%%%%%%%%%%%%%%%%%%%%%
% choose a style
%\bibliographystyle{ieeetr}
%\bibliographystyle{unsrt}
\bibliographystyle{apsrev}
%%%%%%%%%%%%%%%%%%%%%%%%%%%%%%%%%%%%%%%%

%%%%%%%%%%%%%%%%%%%%%%%%%%%%%%%%%%%%%%%%
% choose a .bib file
\bibliography{Bibli}
%%%%%%%%%%%%%%%%%%%%%%%%%%%%%%%%%%%%%%%%

%%%%%%%%%%%%%%%%%%%%%%%%%%%%%%%%%%%%%%%%%%%%%%%%%%%%%%%%%%%%%%%%%%%
% Acknowledgment
\end{document}